\documentclass[%
reprint,
onecolumn,
showkeys,
 amsmath,amssymb,
 aps,
]{revtex4-2}

\usepackage[english]{babel}

\usepackage{graphicx}
\usepackage{dcolumn}
\usepackage{bm}
\usepackage{hyperref}
\hypersetup{
    colorlinks=true,
    linkcolor=blue, 
    urlcolor=blue, 
    citecolor=blue 
}

\newcommand{\transp}{\text{T}}
\newcommand{\partialt}[1]{\partial_t #1}

\newcommand{\partialxx}[1]{\partial^2_x #1}

\newcommand{\mycolon}{\,\text{:}\,}
\newcommand{\mat}[1]{\mbox{\boldmath{$\mathrm{#1}$}}}
\renewcommand{\vec}[1]{\mbox{\boldmath{$#1$}}}
\newcommand{\bkwvec}{\vec{\varphi}^-}
\newcommand{\fwdvec}{\vec{\varphi}^+}

\newcommand{\bkwmat}{\mat{\Phi}^-} 
\newcommand{\fwdmat}{\mat{\Phi}^+}
\newcommand{\clvecfwd}{\vec{\gamma}}
\newcommand{\clmatfwd}{\mat{\Gamma}}

\newcommand{\tqr}{T_\text{QR}}
\newcommand{\spanspc}{\mathop{\mathrm{span}}}

\newcommand{\fbas}{F}
\newcommand{\xbas}{X}

\newcommand{\xvar}{x}
\newcommand{\yvar}{y}

\newcommand{\jacob}{\mat J}

\newcommand{\ramuno}{\mathrm{i}\mkern1mu}

\begin{document}


\title{An improved approach for estimating the dimension of inertial manifolds
  in chaotic distributed dynamical systems via analysis of angles between tangent
  subspaces}

\author{Pavel V. Kuptsov}%
\email{kupav@mail.ru}%
\altaffiliation[Also at ]{Kotelnikov Institute of Radio-Engineering and
  Electronics of Russian Academy of Sciences, Saratov Branch, 38 Zelenaya str.,
  Saratov 410019, Russia}%
\affiliation{ HSE University,\\ 25/12 Bolshaya Pecherskaya str., Nizhny Novgorod
  603155, Russia}%

\date{\today}

\begin{abstract}
  While a previously proposed method for estimating inertial manifold dimension,
  based on explicitly computing angles between pairs of covariant Lyapunov
  vectors (CLVs), employs efficient algorithms, it remains computationally
  demanding due to its substantial resource requirements. In this work, we
  introduce an improved method to determine this dimension by analyzing the
  angles between tangent subspaces spanned by the CLVs. This approach builds
  upon a fast numerical technique for assessing chaotic dynamics
  hyperbolicity. Crucially, the proposed method requires significantly less
  computational effort and minimizes memory usage by eliminating the need for
  explicit CLV computation. We test our method on two canonical systems: the
  complex Ginzburg-Landau equation and a diffusively coupled chain of Lorenz
  oscillators. For the former, the results confirm the accuracy of the new
  approach by matching prior dimension estimates. For the latter, the analysis
  demonstrates the absence of a low-dimensional inertial manifold, highlighting
  a complex regime that merits further investigation. The presented method
  offers a practical and efficient tool for characterizing high-dimensional
  attractors in extended dynamical systems.
\end{abstract}



\maketitle

\section*{Introduction}

While nonlinear dissipative distributed systems possess infinite degrees of
freedom in a formal mathematical sense, their chaotic dynamics often become
confined to a finite-dimensional subspace known as the inertial
manifold. Inertial manifolds possess several important properties: they enclose
the global attractor, exhibit exponential attraction of all trajectories, and
demonstrate stability under perturbations. What is more important, in
infinite-dimensional systems, they enable the reduction of the governing
dynamics to a finite-dimensional set of ordinary differential
equations~\cite{FoiasInManif,RobinsonInManif,Zelik_2014}.

A key obstacle in the study of inertial manifolds is their constructive
description. Despite proven existence for specific systems (e.g.,
Kuramoto-Sivashinsky, complex Ginzburg-Landau~\cite{Temam}), and the
availability of rigorous dimensional bounds from theory, moving from abstract
existence proofs to a concrete geometric understanding is an ongoing challenge.

Numerical work on covariant Lyapunov vectors~\cite{GinCLV,WolfCLV,CLV2012} in
chaotic flows has revealed a fundamental structure~\cite{EffDim,HypDecoup}: in
generic spatially extended dissipative systems, the tangent space splits
hyperbolically. The long-time dynamics are likely confined to a
finite-dimensional subspace of physical Lyapunov modes, physical manifold,
decoupled from an infinite-dimensional subspace of transient Lyapunov modes. As
a basis for the Oseledets subspaces~\cite{Oseled,EckRuell85}, covariant Lyapunov
vectors define the local directions of growth and contraction on the physical
manifold. The vectors associated with the physical manifold display entangled,
frequently tangent dynamics. Conversely, the transient modes are highly damped
and isolated, exhibiting no tangencies or coupling with the entangled
set. Research in~\cite{EffDim,HypDecoup} conjectured that the physical manifold
locally approximates the inertial manifold linearly at every point on the
attractor. According to this hypothesis, the dimension of the inertial manifold
equals the number of entangled Lyapunov modes. This conjecture gained further
support from reference~\cite{GeomInertManif}, which demonstrated that
displacement vectors between recurrent points, i.e., trajectory points that are
temporally distant but spatially close, lie within the local tangent space of
the physical manifold.

The paper~\cite{InertManifUPO} advances this research by showing how to embed
the finite-dimensional physical manifold into the full state space, offering a
path for its explicit construction. The central concept is to use an infinite
set of unstable invariant solutions (e.g., periodic orbits) as a structural
skeleton. This skeleton, combined with locally linear approximations of the
dynamics, provides a complete description of the physical manifold. Within this
framework, chaotic trajectories can be interpreted as a walk across the inertial
manifold, guided by the nearby unstable solutions embedded within it.

The method for estimating the dimension of the inertial manifold, proposed
in~\cite{EffDim,HypDecoup}, is based on the direct computation of covariant
Lyapunov vectors and the angles between them. Although the efficient algorithms
are used for this~\cite{GinCLV,WolfCLV,CLV2012}, the explicit computation of the
vectors and their subsequent pairwise products requires substantial
computational resources. In this work, we propose an improved method for
determining the dimension of the inertial manifold by analyzing the angles
between tangent subspaces spanned by the covariant Lyapunov vectors. This
approach builds upon the fast numerical method for assessing the hyperbolicity
of chaotic dynamics proposed in~\cite{FastHyp12}. In contrast to the method
in~\cite{EffDim,HypDecoup}, the proposed approach demands substantially less
computation and minimizes memory usage by avoiding the need to explicitly
compute covariant Lyapunov vectors.

\section{The Method}\label{sec:method}

Consider a system governed by a set of $m$ ordinary differential equations,
$\dot\xbas=\fbas(t,\xbas)$. The variational equation describing the evolution of
infinitesimal perturbations near a reference orbit $\xbas(t)$ is
\begin{equation}
  \label{eq:linear_common}
  \dot\xvar=\jacob(t)\xvar,
\end{equation}
where $\xbas,\xvar\in \mathbb{R}^m$ represent the state and perturbation
vectors, respectively, and $\jacob(t)\in \mathbb{R}^{m\times m}$ denotes the
Jacobian matrix. The notation $\jacob(t)$ signifies that this matrix depends on
both time $t$ and the state $\xbas(t)$. When $\fbas$, and consequently $\jacob$,
exhibit an explicit dependence on $t$, the system is nonautonomous.

Our methodology is based on the conventional Lyapunov exponent
computation~\cite{Benettin,Shimada79} and involves the sequence of operations
outlined in~\cite{FastHyp12}. The procedure begins by advancing a set of
perturbation vectors along the system's trajectory. Over each time step
$\tqr=t_{n+1}-t_n$, we cycle between solving the variational
equation~\eqref{eq:linear_common} and applying orthonormalization through
Gram–Schmidt or QR decomposition. The choice of the step size $\tqr$ is
arbitrary, provided it is not so large as to cause overflows or underflows.
Following a sufficiently long computation, the perturbation vectors
asymptotically approach the orthonormal backward Lyapunov vectors
$\bkwvec_i(t_n)$ (the Gram-Schmidt vectors), so named because their calculation
is initialized from a remote past state. Proceeding with the algorithm, one
obtains these backward vectors at successive trajectory points indexed by $t_n$.
Each vector $\bkwvec_i(t)$ is associated with its respective Lyapunov exponent
$\lambda_i$, where the exponents follow the conventional descending order. It
is not necessary to execute this procedure using the complete set of $m$
vectors. If the iterations are carried out with a reduced set of $k < m$
perturbation vectors, the result will be limited to the first $k$ backward
Lyapunov vectors and Lyapunov exponents. The orthogonal matrix of these vectors
will be denoted as $\bkwmat_k(t_n)=[\bkwvec_1(t_n),\ldots,\bkwvec_k(t_n)]$

Similarly, the forward Lyapunov vectors $\fwdvec_i(t_n)$ can be computed by
integrating the system forward to a distant future state and then iterating
backward in time from $t_{n+1}$ to $t_n$. The key point here is to perform
backward steps with perturbation vectors using the adjoint variational
equation~\cite{CLV2012,HypDelay2016,HypManyDelay2018}
\begin{equation}
  \label{eq:adjoint_common}
  \dot\yvar=-\jacob^*(t)\yvar.
\end{equation}
Here $\jacob^*(t)$ is the adjoint matrix for $\jacob(t)$, such that the inner
products involving arbitrary vectors $a$ and $b$ satisfy the identity
$\langle \jacob^* a,b\rangle\equiv \langle a,\mat J b\rangle$. If the inner
product is defined as $\langle a,b\rangle=b^\transp a$, as we usually does, then
we have simply $\jacob^*=\jacob^\transp$, where ``$\transp$'' stands for
transposition.

The orthogonal matrices obtained from the QR procedure in the course of the
computations with the adjoint equation~\eqref{eq:adjoint_common} in backward
time converge to the so-called forward Lyapunov vectors~\cite{CLV2012} that can
be collected as an orthogonal matrix
$\fwdmat_k(t_n)=[\fwdvec_1(t_n),\ldots,\fwdvec_k(t_n)]$. By employing the
adjoint equation~\eqref{eq:adjoint_common} instead of the direct
one~\eqref{eq:linear_common}, the forward Lyapunov vectors are obtained in the
correct order: the first vector $\fwdvec_1(t_n)$, corresponding to the largest
Lyapunov exponent $\lambda_1$, emerges first. Conversely, if
Eq.~\eqref{eq:linear_common} were used for the backward iteration, the vectors
would appear in reverse order; that is, the last vector $\fwdvec_m(t_n)$ would
appear first.

Let $\clmatfwd_k(t)=[\clvecfwd_1(t),\clvecfwd_2(t),\ldots,\clvecfwd_k]$ be a
matrix of the first $k$ covariant Lyapunov vectors $\clvecfwd_i(t)$. These
vectors are related to the forward and backward Lyapunov vectors
as~\cite{CLV2012}
\begin{equation}\label{eq:clv_defin}
  \clmatfwd_k(t)=\bkwmat_k(t)\mat{A}_k^-(t)=\fwdmat_k(t)\mat{A}_k^+(t),
\end{equation}
where $\mat{A}_k^-(t)$ is an upper triangular matrix, and $\mat{A}_k^+(t)$ is a
lower triangular matrix. In fact $\bkwmat_k(t)\mat{A}_k^-(t)$ is the QR
decomposition of $\clmatfwd_k(t)$ while $\fwdmat_k(t)\mat{A}_k^+(t)$ is its QL
decomposition~\cite{GolubLoan}. Given the matrices $\bkwmat_k(t)$ and
$\fwdmat_k(t)$ one can compute the CLVs using the equation
$\mat{P}_k(t)\mat{A}_k^-(t)=\mat{A}_k^+(t)$, where
\begin{equation}\label{eq:def_kat_p}
  \mat{P}_k(t)=[\fwdmat_k(t)]^\transp\bkwmat_k(t)
\end{equation}
is an $k\times k$ orthogonal matrix, and $\mat{A}_k^\pm(t)$ are components of
the LU decomposition of $\mat{P}_k(t)$. Details on this method, as well as other
computational procedures for covariant Lyapunov vectors, are provided
in~\cite{CLV2012}. Note that with these algorithms, one can compute any required
number of covariant Lyapunov vectors $1\leq k \leq m$.

The explicit computation of individual covariant Lyapunov vectors can be
avoided. Instead, consider the two tangent subspaces defined as
$S_k=\spanspc\{\clvecfwd_i(t) \mid i=1,\ldots,k\}$ and
$C_k=\spanspc\{\clvecfwd_i(t) \mid i=k+1,\ldots,m\}$. Rather than calculating
all pairwise angles between their constituent vectors, it is sufficient to
compute the smallest principal angle between the subspaces $S_k$ and $C_k$. This
angle vanishes when linear combinations of vectors from the two subspaces
align. Therefore, by evaluating a sequence of these subspace angles for
$k=1,2,\ldots$, one can extract the same information about the tangent space
structure as discussed in~\cite{EffDim,HypDecoup}. The vectors $\clvecfwd_i(t)$
for $i=1,\ldots,k$ need not be found explicitly, because the matrix
$\bkwmat_k(t)$ of the first $k$ backward Lyapunov vectors furnishes the same
basis. Furthermore, rather than working with $C_k$ directly, it is more
efficient to use its orthogonal complement $C_k^\bot$, whose basis is provided
by the matrix $\fwdmat_k(t)$.

Consider the matrix $\mat{P}_k(t)$, see~\eqref{eq:def_kat_p}. Its singular
values are the cosines of the principal angles between the subspaces $S_k$ and
$C_k^\bot$ (see Ref.~\cite{GolubLoan} for details on principal angles between
subspaces). Since $C_k^\bot$ is the orthogonal complement to the subspace of
interest $C_k$, the smallest principal angle $\theta_k$ between $S_k$ and $C_k$
is indicated by the smallest singular value $\sigma_k$ of $\mat{P}_k(t)$:
\begin{equation}
  \label{eq:prin_angle}
  \theta_k=\pi/2-\arccos \sigma_k.
\end{equation}
See Refs.~\cite{CLV2012} for the mathematical details.

Assume we want to check the angles for subspaces $S_k$ and $C_k$ for
$k=1,2,\ldots n$, where $n\leq m$. First, we compute the bases $\bkwmat_n(t)$
and $\fwdmat_n(t)$ as described. Next, we compute the matrix $\mat{P}_n(t)$
(see~\eqref{eq:def_kat_p}) and consider the sequence of its upper-left
submatrices $\mat P(1\mycolon k,1\mycolon k)$ for $k=1,2,\ldots, n$. The desired
sequence of angles between successive tangent subspaces is then computed
according to Eq.~\eqref{eq:prin_angle} using the singular values of
$\mat P(1\mycolon k,1\mycolon k)$.

Note that a similar idea underlies the hyperbolicity and pseudohyperbolicity
test for chaotic systems, which is based on verifying an angle
criterion~\cite{FastHyp12,CLV2012,HypDelay2016,HypManyDelay2018,PsHyp2018}.

\section{Inertial manifold dimensions for nonlinear distributed system}

\subsection{Complex Ginzburg-Landau equation}

Consider the one-dimensional complex Ginzburg–Landau equation
\begin{equation}
  \label{eqn:cgle}
  \partialt{u} = u - (b_3+\ramuno c_3) u|u|^2 + (b_1+\ramuno c_1) \partialxx{u}.
\end{equation}
where $u(t,x)$ is complex variable, and $c_1$ and $c_3$ are real control
parameters. For numerical computation of its solutions, the second spatial
derivative will be approximated using a finite-difference scheme. This
transforms the partial differential equation into a system of $N$ coupled
Landau–Stuart oscillators:
\begin{equation}
  \label{eqn:cgle_chain}
  \dot{u}_n = u_n - (b_3+\ramuno c_3)u_n|a_n|^2 + (b_1+\ramuno c_1)\kappa(u_n)/h^2,
\end{equation}
Now $u_n\equiv u_n(t)$ ($n=0,1,\dots,N-1$), where $N$ is a number of
discretization steps of a spatial area of length $L$. Function $\kappa(u_n)$
determines the diffusive coupling and no-flux boundary conditions:
\begin{equation}
  \label{eq:coupling}
  \begin{aligned}
    \kappa(u_n) &= u_{n-1}-2u_n+u_{n+1}, \; n=1,2,\ldots N-2,\\
    \kappa(u_0) &= 2(u_1-u_0), \\
    \kappa(u_{N-1}) &= 2(u_{N-2}-u_{N-1}).
  \end{aligned}
\end{equation}
Step size of the discretization $h$ is defined as $h=L/(N-1)$. As $N$ grows for
constant $L$, the system approaches the continuum limit, allowing us to draw
conclusions about the properties of the original equation~\eqref{eqn:cgle}.

At $c_3=3$, $c_1=-2$, $b_3=1$, $b_1=1$ the system under consideration
demonstrates the regime of so called ``amplitude
turbulence''~\cite{CgleChaos}. Lyapunov exponents for this regime are plotted in
Fig.~\ref{fig:cgle_lyap1}. The Lyapunov exponents are computed for increasing
$N$ at constant $L$ and one sees that the curves are almost identical. Thus one
can expect that they will remain the same in the continuity limit so that this
spectrum represent the true properties of the original continues
system~\eqref{eqn:cgle}. The first five values of them at $N=40$ and $L=8$ are
$\lambda_{1,\dots,5}=1.20\times 10^{-1}, 3.17\times 10^{-4}, -5.62\times
10^{-4}, -5.21\times 10^{-2}, -3.17\times 10^{-1}$. The values
$3.17\times 10^{-4}$, and $-5.62\times 10^{-4}$ should be treated as numerical
zeros: zero Lyapunov exponents is always present for autonomous flow systems and
the other one is related to the symmetry of Eq.~\eqref{eqn:cgle_chain}. The
corresponding Kaplan-Yorke dimension is $D_{\text{KY}}=4.21$. Note that this
value is much smaller compared with the dimension of the phase space which in
this case equals to $2N=80$. It indicates that the dynamics is indeed confined
confined to a low-dimensional subspace.

\begin{figure}[!ht]
  \centering\includegraphics{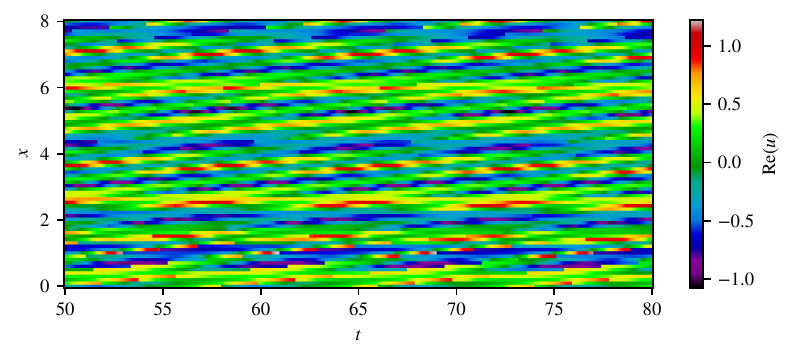}
  \caption{Amplitude turbulence in the system~\eqref{eqn:cgle_chain}. $L=8$,
    $N=80$, $c_3=3$, $c_1=-2$, $b_3=1$, $b_1=1$} \label{fig:cgle_sptm1}
\end{figure}

\begin{figure}[!ht]
  \centering\includegraphics{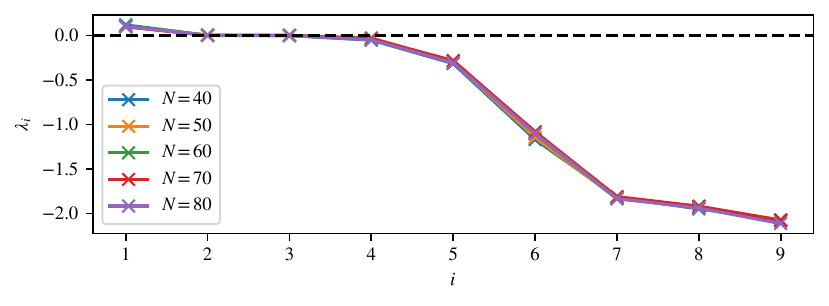}
  \caption{Lyapunov exponents of the system~\eqref{eqn:cgle_chain} at the
    parameters as in Fig.~\ref{fig:cgle_sptm1} for various
    $N$.} \label{fig:cgle_lyap1}
\end{figure}

\begin{figure}[!ht]
  \centering\includegraphics{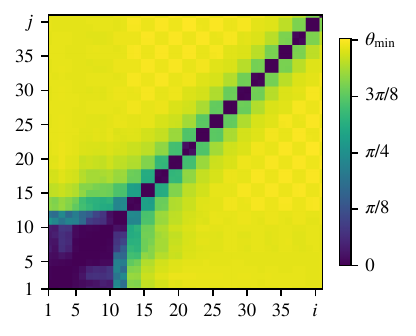}
  \caption{The pairwise angles between covariant Lyapunov vectors for the
    system~\eqref{eqn:cgle_chain} at $N=40$. All other parameters are as in
    Fig.~\ref{fig:cgle_sptm1}} \label{fig:cgle_angpair}
\end{figure}

Figure~\ref{fig:cgle_angpair} shows the pairwise angles between covariant
Lyapunov vectors for the system~\eqref{eqn:cgle_chain} at $N=40$ computed to
reproduce the method described in~\cite{EffDim,HypDecoup}. It is evident that
the first ten vectors are strongly entangled among themselves, while being
decoupled from all subsequent vectors. Therefore, the dimension of the inertial
manifold in this case can be estimated as 10. Note that although this value is
much smaller than the full phase space dimension $2N=80$, it is nevertheless
higher than $D_{\text{KY}}$. This indicates that the dimension of the attractor
significantly underestimates the dimension of the inertial manifold.

\begin{figure}[!ht]
  \centering\includegraphics{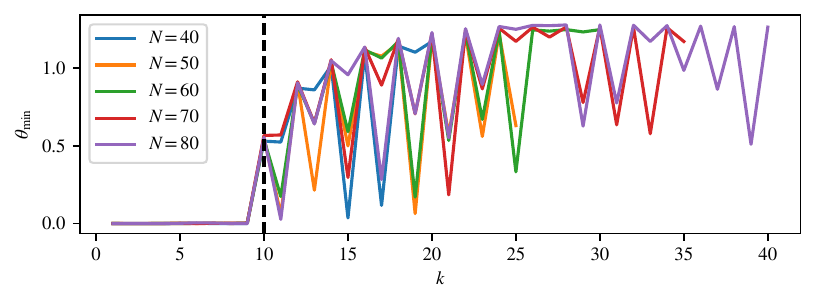}
  \caption{Angles $\theta_k$ between tangent subspaces $S_k$ and $C_k$. The
    first non-vanishing angle at $k=10$ (marked as vertical dashed line)
    indicates the dimension of the inertial manifold. Observe that it is the
    same for all $N$.} \label{fig:cgle_angles1}
\end{figure}

Figure~\ref{fig:cgle_angles1} demonstrates computation of inertial manifold
dimension by using our methods, described in Sec.~\ref{sec:method}. The
dimension is indicated by the first non-vanishing angle at $k=10$. Note that
this is equal to the estimate obtained above via pairwise angles. Also note that
it remains the same when $N$ increases and it means that 10 is a good estimate
for the inertial manifold of the continues system~\eqref{eqn:cgle}.

\begin{figure}[!ht]
  \centering\includegraphics{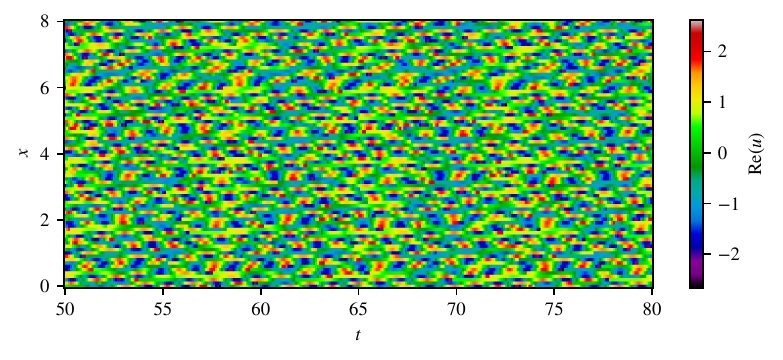}
  \caption{Spatiotemporal dynamics of the system~\eqref{eqn:cgle_chain} at
    $c_3=3$, $c_1=0$, $b_3=1$, $b_1=-3.0$. $N=80$, $L=8$.} \label{fig:cgle_sptm2}
\end{figure}

\begin{figure}[!ht]
  \centering\includegraphics{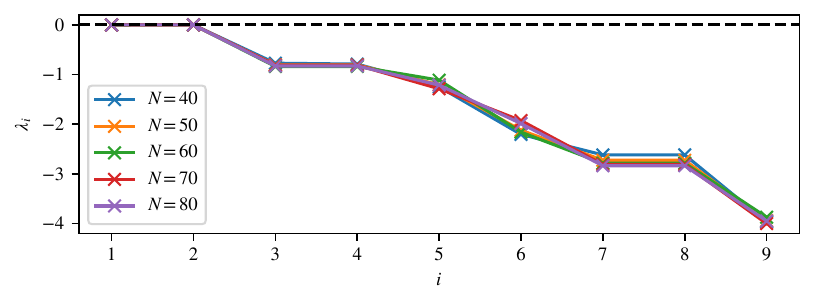}
  \caption{Lyapunov exponents of the system~\eqref{eqn:cgle_chain}. Parameters
    are as in Fig.~\ref{fig:cgle_sptm2}.} \label{fig:cgle_lyap2}
\end{figure}

One more example is computed for $c_3=3$, $c_1=0$, $b_3=1$, $b_1=-3.0$. The
spatiotemporal behavior is shown in Fig.~\ref{fig:cgle_sptm2}. Although it
appears irregular, the system is actually in a quasiperiodic regime, as can be
concluded from the Lyapunov exponent spectrum in Fig.~\ref{fig:cgle_lyap2}. One
can see that all Lyapunov exponents in this case are non-positive. For example
at $N=40$ the first five exponents are as follows:
$\lambda_{1,\ldots,5}=-1.53\times 10^{-7}, 3.07\times 10^{-6}, -7.72\times
10^{-1}, -7.90\times 10^{-1}, -1.25$. As above the two values
$-1.53\times 10^{-7}$, $3.07\times 10^{-6}$ should be treated as numerical
zeros. Note again that the Lyapunov spectra computed for different $N$ at
constant $L$ coincide almost perfectly which indicates that they remain the same
in the continuity limit.

\begin{figure}[!ht]
  \centering\includegraphics{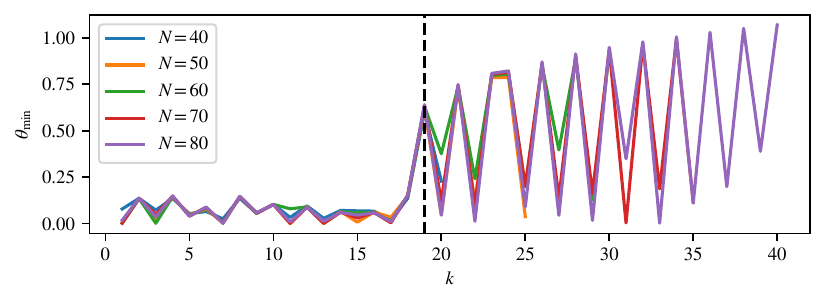}
  \caption{} \label{fig:cgle_angles2}
\end{figure}

The angles $\theta_k$ for this case are shown in Fig.~\ref{fig:cgle_angles2}. It
can be observed that there is no region of vanishing angles, which, as discussed
above, indicates the presence of an inertial manifold. However, clear remnants
of such a region are still visible. The angle $\theta_k$ fluctuates near zero in
the left portions of the curves for all values of $N$, but its behavior changes
markedly at $k=19$. Thus, while no inertial manifold is revealed here in a
rigorous sense, one can hypothesize the existence of a manifold with special
properties—a sort of ``weak inertial manifold''.

Thus we observe that the Ginzburg-Landau equation does not always possess a
well-defined inertial manifold. Further discussion of this can be found
in~\cite{StrictFussy}.

\subsection{Chain of Lorenz systems}

Now we consider another example of a distributed system with chaotic dynamics: a
chain of coupled Lorenz systems. We incorporate diffusive coupling via the $y$
variables as follows:
\begin{equation}
  \begin{aligned}
    \dot x_n &= \sigma(y_n-x_n)\\
    \dot y_n &= r x_n - y_n - x_n z_n + \epsilon (y_{n-1}-2y_n+y_{n+1})\\
    \dot z_n &= x_ny_n - b z_n
  \end{aligned}\label{eq:lorenz_chain}
\end{equation}
where $\sigma=10$, $r=28$, $b=8/3$ are standard values of the control
parameters, $\epsilon$ is the coupling strength and no-flux boundary conditions
are employed, see Eq.~\eqref{eq:coupling}.

A single Lorenz system is known to possess a pseudohyperbolic
attractor~\cite{TurShil98,TurShil08,PsReview}. Such attractors are true strange
attractors, as every orbit exhibits a positive Lyapunov exponent; that is, no
stable periodic orbits exist. This characteristic is robust and persists under
sufficiently small perturbations. In pseudohyperbolic systems, the tangent space
splits into a direct sum of volume-expanding and volume-contracting
subspaces. This splitting remains invariant over time, and the subspaces cannot
become tangent to one another. For the Lorenz attractor the dimension of the
volume-expanding subspace is 2 and the volume-contracting is
1-dimensional. Numerical verification of these properties is provided
in~\cite{PsHyp2018}.

When the coupling parameter $\epsilon$ is sufficiently small, an entire chain of
$N$ Lorenz systems~\eqref{eqn:cgle_chain} also exhibits pseudohyperbolic
properties due to the robustness of the individual Lorenz attractors. In such a
chain, the expanding and contracting subspaces are $2N$- and $N$-dimensional,
respectively. In this case, no inertial manifold exists because each local
attractor remains nearly unaltered; consequently, they collectively occupy the
entire phase space, and the overall dynamics remains high-dimensional.

This case is illustrated in
Figs.~\ref{fig:lorenz_sptm_0,5},~\ref{fig:lorenz_lyap_0,5} and
\ref{fig:lorenz_ang_0,5}. Spatiotemporal diagram in
Fig.~\ref{fig:lorenz_sptm_0,5} contains some regular patterns, but the
corresponding Lyapunov exponents in Fig.~\ref{fig:lorenz_lyap_0,5} indicates a
chaotic dynamics. Kaplan-Yorke dimension for $N=40, 60, 80$ is
$D_{\text{KY}}=80.4, 120.6, 160.7$, respectively.

\begin{figure}[!ht]
  \centering\includegraphics{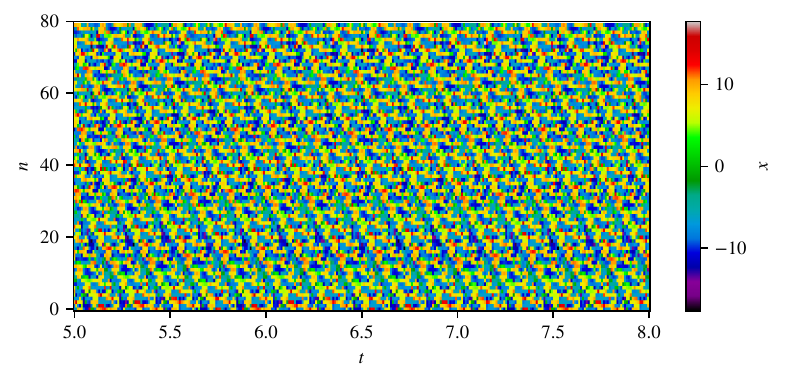}
  \caption{Spatiotemporal dynamics of the chain~\eqref{eq:lorenz_chain} with
    $N=80$ elements at $\epsilon=0.5$} \label{fig:lorenz_sptm_0,5}
\end{figure}

\begin{figure}[!ht]
  \centering\includegraphics{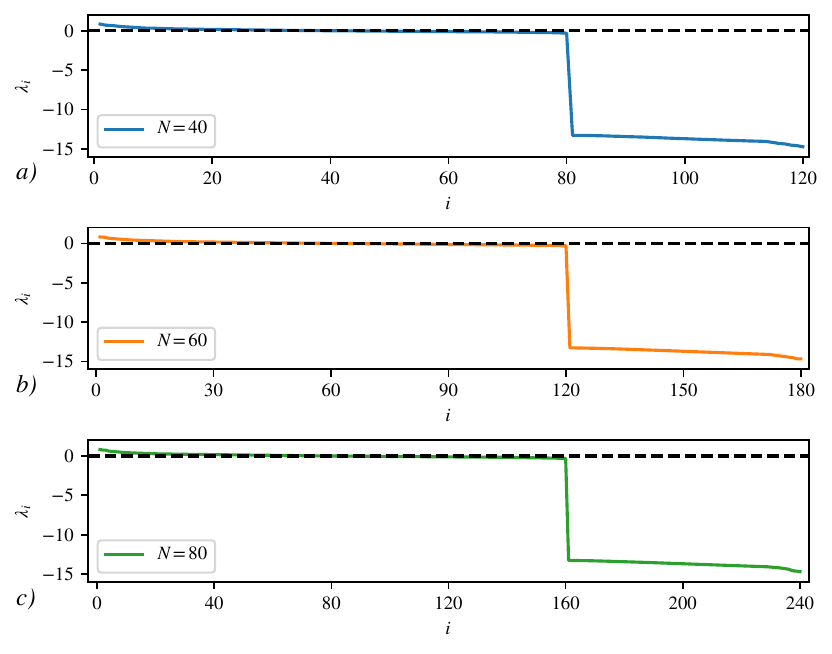}
  \caption{Lyapunov exponents spectra for the chain~\eqref{eq:lorenz_chain} at
    $\epsilon=0.5$. a) $N=40$, b) $N=60$ and c) $N=80$. Observe positive values
    that indicate chaos and the sharp gaps located at $i=2N$.
  }\label{fig:lorenz_lyap_0,5}
\end{figure}

A necessary condition for the existence of the pseudohyperbolic attractor is the
following relation for its Lyapunov exponents~\cite{TurShil98,TurShil08}
\begin{equation}
  \label{eq:pseudo_lyap_cond}
  \sum_{i=1}^K\lambda_i>0, \text{ and }\lambda_i<0 \text{ for }i>K.
\end{equation}
Here $K$ is the dimension of the volume-expanding tangent subspace of the
pseudohyperbolic attractor. 

Testing the condition~\eqref{eq:pseudo_lyap_cond} for the computed numerical
values of the Lyapunov exponents plotted in Fig.~\ref{fig:lorenz_lyap_0,5} we
obtain that the condition $K=2N$ fulfills remarkably exact for all tested values
of $N$.

The angles between the tangent subspaces are displayed in
Fig.~\ref{fig:lorenz_ang_0,5}. The features typically observed when an inertial
manifold exists are not visible here: there are no clear regions of vanishing
versus non-vanishing angles. The nonzero angle at $k=2N$ indicates the absence
of tangencies between the $2N$-dimensional volume-expanding and the
$N$-dimensional volume-contracting subspaces of the chain. Together with the
fulfillment of condition~\eqref{eq:pseudo_lyap_cond} also at $k=2N$, this
confirms that for small $\epsilon$, the chain possesses no inertial manifold and
its phase space is instead filled by a high-dimensional pseudohyperbolic
attractor.

\begin{figure}[!ht]
  \centering\includegraphics{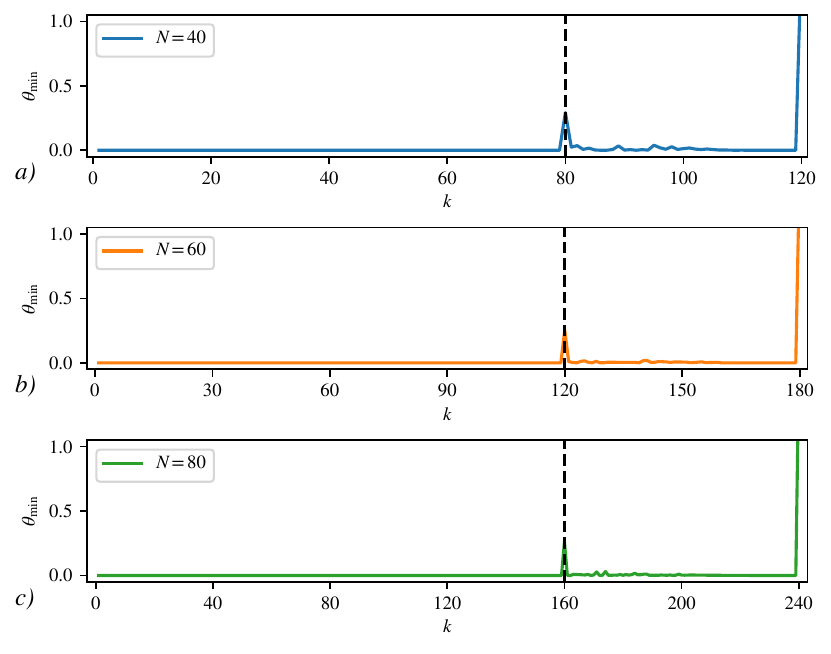}
  \caption{Angles between tangent subspaces $\theta_k$ for the
    chain~\eqref{eq:lorenz_chain} at $\epsilon=0.5$. Panels a), b) and c)
    correspond to $N=40$, $N=60$, and $N=80$, respectively. The dashed vertical
    lines are plotted at $k=2N$ where the nonzero angle indicate the absence of
    tendencies between $2N$-dimensional volume-expanding and $N$-dimensional
    volume-contracting subspaces of the chain.} \label{fig:lorenz_ang_0,5}
\end{figure}

\begin{figure}[!ht]
  \centering\includegraphics{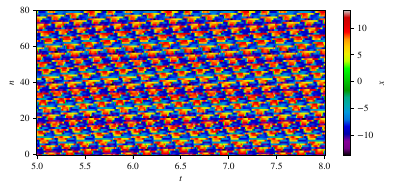}
  \caption{Spatiotemporal dynamics of the chain~\eqref{eq:lorenz_chain} with
    $N=80$ at $\epsilon=5$} \label{fig:lorenz_sptm_5}
\end{figure}

\begin{figure}[!ht]
  \centering\includegraphics{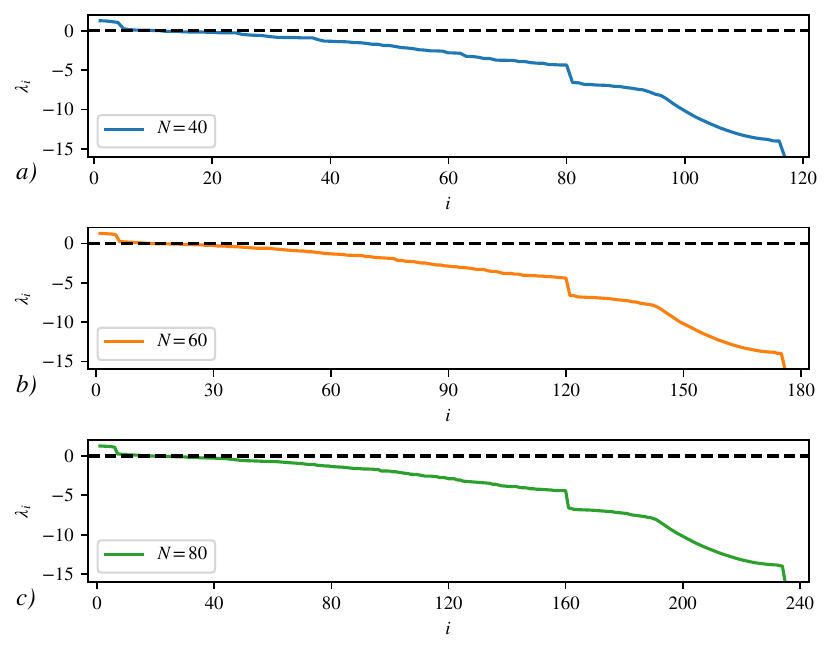}
  \caption{Lyapunov spectra for the chain at $\epsilon=5$. Observe small number
    of positive values of the exponents.} \label{fig:lorenz_lyap_5}
\end{figure}

\begin{figure}[!ht]
  \centering\includegraphics{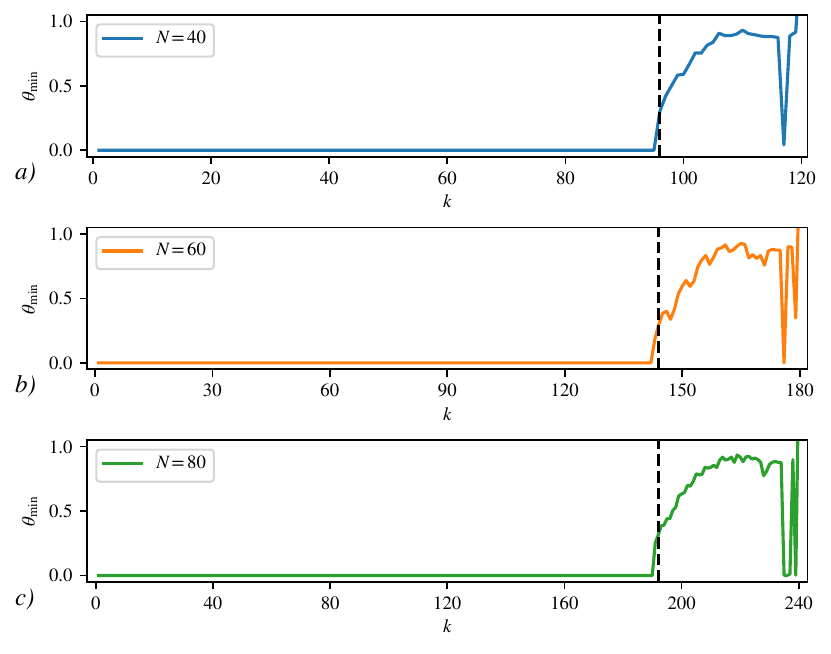}
  \caption{The angles $\theta_k$ between tangent subspaces
    of~\eqref{eq:lorenz_chain} at $\epsilon=5$. The dashed vertical line marks
    the boundary separating the vanishing and non-vanishing angles at
    $k=0.8 \times 3N$: a) $k=96$, b) $k=144$, c)
    $k=192$.} \label{fig:lorenz_ang_5}
\end{figure}

Now we consider the case of strong coupling $\epsilon=5$, see
Figs.~\ref{fig:lorenz_sptm_5},~\ref{fig:lorenz_lyap_5} and
\ref{fig:lorenz_ang_5}. Spatiotemporal diagram in Fig.~\ref{fig:lorenz_sptm_5}
together with Lyapunov spectra in Fig.~\ref{fig:lorenz_lyap_5} again indicates
chaotic dynamics of the chain. But the number of positive exponents is
small and Kaplan-Yorke dimensions $D_{\text{KY}}=10.8, 21.4, 30.1$ for
$N=40, 60, 80$, respectively, are significantly smaller in comparison with the
previous case $\epsilon=0.5$

The spatiotemporal diagram in Fig.~\ref{fig:lorenz_sptm_5}, along with the
Lyapunov spectra shown in Fig.~\ref{fig:lorenz_lyap_5}, again confirms the
presence of chaotic dynamics in the chain. However, the number of positive
exponents is small, and the Kaplan-Yorke dimensions --- specifically
$D_{\text{KY}}=10.8, 21.4, 30.1$ for $N=40, 60, 80$, respectively --- are
significantly lower compared to the previous case with $\epsilon=0.5$.

The curve of angles between tangent subspaces $\theta_k$ demonstrates two
expected areas of vanishing and non-vanishing values. However, the point where
it occurs $k=0.8 \times 3N$ indicates very large dimension for a candidate for
an inertial manifold. This splitting point lies even beyond the point $k=2N$
observed for a high-dimensional attractor in the previous case. Thus we can
conclude that although the fractal dimension of the attractor estimated via
Kaplan-Yorke dimension is relatively small, a low dimensional inertial manifold
in this case does not exist. In summary, the attractor of the system
\eqref{eq:lorenz_chain} at various coupling strength merits more detailed
analysis.

\section*{Conclusion}

In this paper, we propose an approach for estimating the dimension of an
inertial manifold in extended systems. The method is based on the hypothesis
that the inertial manifold can be identified by examining the angles between the
system's tangent subspaces. Compared to the previously suggested technique,
which relies on pairwise angles between covariant Lyapunov vectors, the
presented approach is more computationally efficient, as it avoids the need for
explicitly computing these covariant vectors.

We demonstrate the proposed method using two examples: the complex
Ginzburg-Landau equation and a diffusely coupled chain of Lorenz
oscillators. For the Ginzburg-Landau equation, the computed dimension of the
inertial manifold agrees with the result obtained using the previous method,
which relied on the explicit use of covariant Lyapunov vectors. For the chain of
Lorenz systems, we show that this system possess a high-dimensional
pseudohyperbolic attractor and a low-dimensional inertial manifold does not
exist, a finding which warrants further detailed analysis.

\begin{acknowledgments}
This work was supported by the Russian Science Foundation, Russia,
25-11-20069, \url{https://rscf.ru/en/project/25-11-20069}
\end{acknowledgments}

\bibliographystyle{unsrt} 
\bibliography{inert_manif}

\end{document}